\documentclass[pra,twocolumn,superscriptaddress,showpacs]{revtex4-1}
\usepackage{amssymb,amsmath,graphicx,color}
\usepackage{times}

\begin{document}

\title{Topological quantum memory interfacing atomic and superconducting qubits}

\author{Zheng-Yuan Xue}
\affiliation{Guangdong Provincial Key Laboratory of Quantum Engineering and Quantum Materials, and School of Physics\\ and Telecommunication Engineering,
South China Normal University, Guangzhou 510006, China}

\author{Zhang-qi Yin} \email{yinzhangqi@gmail.com}
\affiliation{Center for Quantum Information, Institute of Interdisciplinary Information Science, Tsinghua University, Beijing 100084, China}

\author{Yan Chen}
\affiliation{Department of Physics, State Key Laboratory of Surface Physics and
Laboratory of Advanced Materials,\\ Fudan University, Shanghai 200433, China}

\author{Z. D. Wang} \email{zwang@hku.hk}
\affiliation{Department of Physics and Center of Theoretical and Computational Physics, The University of Hong Kong,\\ Pokfulam Road, Hong Kong, China}

\author{Shi-Liang Zhu} \email{slzhu@nju.edu.cn}
\affiliation{National Laboratory of Solid State Microstructure and Department of Physics, Nanjing University, Nanjing 210093, China}

\date{\today}

\begin{abstract}
We propose a scheme to manipulate a topological spin qubit which is realized with cold atoms in a one-dimensional optical lattice.  In particular, by introducing a quantum opto-electro-mechanical interface, we are able to first transfer a superconducting qubit state to an atomic qubit state and then to store it into the  topological spin qubit. In this way, an efficient  topological quantum memory could be constructed for the superconducting qubit. Therefore, we can consolidate the advantages of both the noise resistance of the topological qubits and the scalability of the superconducting qubits in this hybrid architecture.
\end{abstract}


\maketitle

\section{Introduction}

Quantum computation has attracted much attention as it is
able to solve diverse classes of
hard problems. Superconducting circuits are
promising for  implementing quantum computer hardwares as they
are potentially scalable \cite{s2}.  As
a superconducting qubit is usually quite sensitive to the external
environments and background noises, its coherence time is
generally rather short \cite{s1}.
A promising strategy out of this difficulty is based
on the topological idea \cite{k03}: a topological qubit is
largely insensitive to major sources of  local noises, and thus can be
used to form an efficient topological quantum memory (TQM).

Recently, with the potential applications in topological quantum
computation, topological states of matter  have attracted   renewed
interests \cite{p,mz,1d}. In particular, time-reversal  invariant
topological insulators \cite{t1,t2,t3,t4,t5} have been reported
experimentally, and thus have greatly stimulated  the study of
topological phases \cite{mz}. In engineering topological phases, the
spin-orbit (SO) interaction usually plays an important role.
Therefore, with recent great achievements in realizing artificial
SO interaction in cold atom system
\cite{gauge1,gauge2,gauge3,Lin,Wang,MIT,Pan}, it becomes a new
platform  to probe   topological phases  in a fully controllable
way  \cite{simulation}. Recently, Liu \emph{et al}. \cite{liu}
proposed to observe and manipulate topological edge spins realized
in a one-dimensional (1D) optical lattice with experimentally
realized SO interaction. The nontrivial topology there supports
two degenerate  zero modes, which are topologically stable, and thus
can be used to construct  a topological spin qubit (TSQ).

\begin{figure}[tb]\begin{center}
\includegraphics[width=8cm]{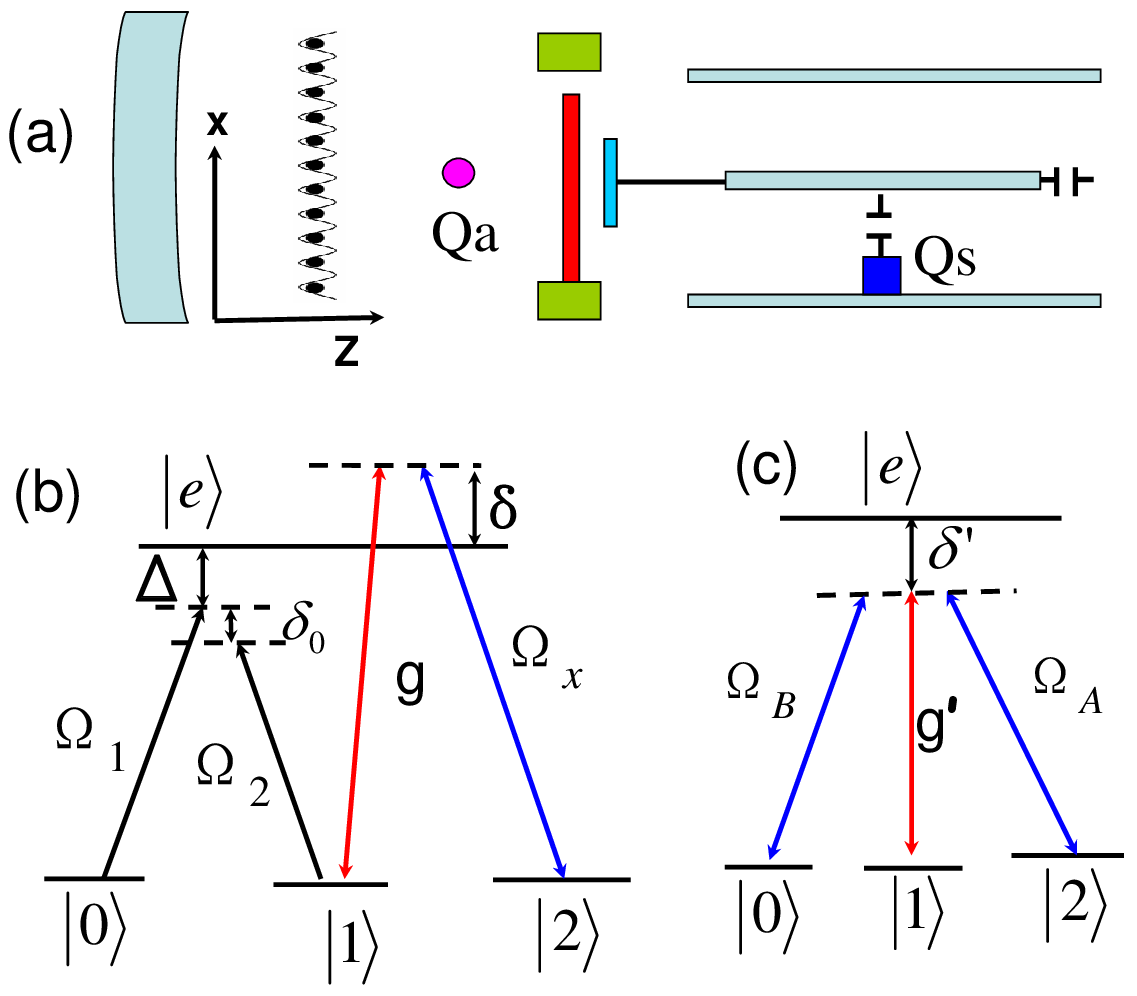} \end{center}
\caption{Illustration of our scheme. (a) A hybrid architecture consists of
two-cavity optomechanical system.  A mechanical oscillator (red)
mediates the coupling of the optical (left) and microwave cavities
(right). A 1D optical lattice, constructing our TQM, is inside the
optical cavity. An ancillary atom (pink circle) is incorporated to
engineer the cavity photon state and its information can be stored
in the TQM. (b) The level structure of atoms in the optical
lattice. The transitions $|0\rangle\rightarrow|e\rangle$ and
$|1\rangle\rightarrow|e\rangle$ are induced by the lasers in a
large one-photon detuning $|\Delta|\gg\Omega$ and a small
two-photon detuning $|\delta_0|\ll\Omega$. The transition
$|1\rangle\rightarrow|e\rangle$ is dispersively coupled to the
cavity field to achieve the QND Hamiltonian. To obtain effective
switch of the  cavity-assisted interaction, a strong control laser
of Rabi frequency  $\Omega_x$ driving the transition of
$|2\rangle\rightarrow|e\rangle$ is also introduced. (c)  The level
structure of the ancillary atom. }  \label{coupled}
\end{figure}

Here, we propose a scheme to realize an interface between
this TSQ and  a solid-state superconducting qubit. This hybrid
system may allow us to combine the advantages of both the noise
resistance of the topological qubits and the scalability of the
superconducting qubits. With the help of cavity assisted
interaction \cite{jiang}, we show that local operations can be
implemented for the TSQ. Our particular interest lies in using
this TSQ as a TQM, where we can store quantum information of both
atomic and superconducting qubits. Meanwhile, recent experiments shown that
the TSQ was not only a theoretical proposal, but also within
experimental feasibility \cite{Wu2015}. Finally, we note that there are proposals considering the hybrid systems consisting of  superconducting qubits and  nitrogen-vacancy centers. However, as  the coupling between a single nitrogen-vacancy center and a cavity is usually very weak, below 0.1 kHz level \cite{h1}, previous works mainly focus on using ensemble of nitrogen-vacancy centers \cite{h1,h2,h3,h4}, where the dephasing time of the ensemble is much shorter comparing with  a single nitrogen-vacancy center.

To store the state of a
superconducting qubit [blue rectangle in Fig. 1(a)] into the TSQ
formed by the atomic lattice, as shown in Fig. 1(a), we  firstly
transfer the state of the superconducting qubit to an ancillary
atomic qubit [pink circle in Fig. 1(a)] and then store it into the
TSQ. As the superconducting and ancillary atomic qubits are of
vastly different frequencies, a quantum opto-electro-mechanical setup
is needed, where a mechanical oscillator mediates the
coherent coupling of both microwave and optical photons
\cite{strong,interface,om,linear,Yin2015}. It was reported that this
interface had been realized in recent experiments \cite{strong,interface}.
The superconducting qubit interacts with a
microwave cavity mode in a circuit QED scenario \cite{cqed}, while the
atoms interact with the optical cavity mode. When  the
intra-cavity interaction is switched on, one can obtain high fidelity state
transfer from a superconducting qubit to the ancillary atomic
qubit. By  switching off the intra-cavity interaction, we show
that a TQM for the ancillary atom can be constructed. Combining
the two processes become particularly arresting as we can store
a   superconducting qubit state into the TQM, which provides an alternation for interfacing topological and superconducting qubits \cite{inter1,inter3,inter2,inter4}.

\section{A topological spin qubit}

The TSQ is based on a  quasi-1D cold fermions with
three-level Lamda configuration trapped in an optical lattice \cite{liu}. As
shown in Fig. 1(b), the transitions $|0\rangle\rightarrow|e\rangle$ and $|1\rangle\rightarrow|e\rangle$ are induced by the lasers, with Rabi-frequencies $\Omega_1=\Omega\sin(k_0x/2)$ and $\Omega_2=\Omega\cos(k_0x/2)$, in a large one-photon detuning $|\Delta|\gg\Omega$ and a small two-photon detuning $|\delta_0|\ll\Omega$, which
is equivalent to a Zeeman field along the $z$ axis
$\Gamma_z=\hbar\delta_0/2$ and can be precisely controlled.
Then, adiabatically eliminating the excited state $|e\rangle$
yields the following effective Hamiltonian
\begin{eqnarray}\label{eqn:H2}
H_{\rm eff}=\frac{p_x^2}{2m}+\sum_{\sigma={0,1}}\bigr[V_\sigma(x)+\Gamma_z\sigma_z\bigr]|\sigma\rangle\langle \sigma|\notag\\
-\bigr[M(x)|0\rangle\langle 1|+{\rm H.c.}\bigr],
\end{eqnarray}
where $M(x)=M_0\sin(k_0x)$ with
$M_0=\frac{\hbar\Omega^2}{2\Delta}$ being an additional
laser-induced Zeeman field along $x$ axis. To form the 1D lattice,
the optical dipole trapping  potentials are chosen as
$V_\sigma(x)=-V_0\cos^2(k_0x)$ with the lattice trapping
frequency being $\omega=(2V_{0}k_0^2/m)^{1/2}$ \cite{simulation}. The states
$|0\rangle$ and $|1\rangle$ are defined by spin up and down of a
pseudo-spin, respectively. The tight-binding description of Hamiltonian (\ref{eqn:H2}) is defined as the case when  the fermions occupy the lowest $s$-orbitals $\phi_{s\sigma}$. Redefining the spin-down operator as $\hat
c_{j\downarrow}\rightarrow e^{i\pi x_j/a}\hat c_{j\downarrow}$
with $a$ being lattice constant,  the tight-binding Hamiltonian
reads \cite{liu}
\begin{eqnarray}\label{eqn:tightbinding2}
H&=&-t_s\sum_{<i,j>}(\hat c_{i\uparrow}^{\dag}\hat
c_{j\uparrow}-\hat c_{i\downarrow}^{\dag}\hat
c_{j\downarrow})+\sum_{i}\Gamma_z(\hat n_{i\uparrow}-\hat n_{i\downarrow})\nonumber\\
&&+\left[\sum_{j}t_{\rm so}^{(0)}\left(\hat c_{j\uparrow}^\dag\hat c_{j+1\downarrow}
-\hat c_{j\uparrow}^\dag\hat c_{j-1\downarrow}\right)+{\rm H.c.}\right],
\end{eqnarray}
where $\hat n_{i\sigma}=\hat c_{i\sigma}^\dag\hat c_{i\sigma}$,
$$t_s=\int
dx\phi^{(j)}_{s\sigma}(x)\left[\frac{p_x^2}{2m}+V^{\rm}\right]\phi^{(j+1)}_{s\sigma}(x),$$
$$t_{\rm so}^{(0)}=\frac{\Omega_0^2}{\Delta}\int
dx\phi_{s}(x)\sin(2k_0x)\phi_{s}(x-a).$$ In  $k$ space, the
tight-binding Hamiltonian can be rewritten as
\begin{equation}
H_k=-\sum_{k,\sigma\sigma'}\hat c_{k,\sigma}^{\dag}
\left[d_z(k)\sigma_z+d_y(k)\sigma_y\right]_{\sigma,\sigma'}\hat
c_{k,\sigma'},
\end{equation}
where $d_y=2t_{\rm so}^{(0)}\sin(ka)$
and $d_z=-\Gamma_z+2t_s\cos(ka)$.
This Hamiltonian describes a nontrivial topological insulator when $|\Gamma_z|<2t_s$ and otherwise a  trivial insulator, with a bulk gap
$E_g=\mbox{min}\{|2t_s-|\Gamma_z||, 2|t_{\rm so}^{(0)}|\}$.
The nontrivial topology supports two degenerate boundary modes, and
each mode equals one-half of a spin 1/2 particle, similar to the
relation between a Majorana fermion and a complex fermion in
topological superconductors \cite{kitaev}. The zero
modes are robust to  local noises, and thus may form  a TSQ which
we can use as a TQM. The TSQ can also be obtained in the
middle of the lattice  by creating
mass domain. Local operations upon the TSQ can be achieved by
applying a local Zeeman term $\textbf{B}_{y}=\Gamma_0\sigma_y$ or
$\textbf{B}_z=\Gamma_0\sigma_z$ \cite{liu}. When
$|\Gamma_0|>2|t_{\rm so}^{(0)}|$ a mass domain is created,
associated with two midgap spin states $|\psi_\pm\rangle$
localized around the two edges of the TSQ.
Let $|\psi_+\rangle$ be initially occupied, reducing $|\Gamma_0|$
smoothly can open the coupling in $|\psi_\pm\rangle$ and lead the
TSQ state to evolve. If we apply the Zeeman field along $z$ ($y$)
axis, the TSQ evolves in the $x$-$y$ ($x$-$z$) plane \cite{liu}.

\section{TQM for an atomic qubit}
We then show how to store the state of an ancillary atom
into the TSQ. For our TQM purpose, we need controlled manipulation
of the TSQ, and thus a controllable local Zeeman field. Here, we
propose to implement this controlled manipulation with
cavity-assisted quantum nondemolition (QND) Hamiltonian
\cite{jiang}. To achieve this, we introduce another level, and
thus the fermions are now in four-level Tripod  configuration. For
$^6$Li and $^{40}$K atoms, we can find appropriate  hyperfine
levels to meet the requirement. The coupling structure is shown
in Fig. 1(b).  To implement the QND Hamiltonian $H_{QND}=\chi
a^\dagger a\sum_l \sigma_l ^z$, we set the cavity mode to couple
the transition of $|1\rangle\leftrightarrow|e\rangle$ with a
strength $g$ and blue detuning $\delta$, where
$\chi=g^2/(2\delta)$ and $a (a^{\dagger})$ is the annihilation
(creation) operator for the cavity photon. As the cavity assisted
interaction is always-on, if we want the interaction
to act only on the atoms within certain area, we should be able to
decouple the interaction outside the wanted area. This can be done
by using another laser which couples the transition of excited
state to another level with a Rabi frequency $\Omega_x$, i.e.,
$|2\rangle\leftrightarrow|e\rangle$,
as shown in Fig. 1(b). When   $\Omega_x\gg g$, the destructive
interference of excitation pathways from the two transition
ensures that the so-called dark state, which decoupled the atoms
from interacting with both optical fields \cite{decouple1,decouple2}. This
QND Hamiltonian preserves the photon number $n_c$ of the cavity
mode. Within the $n_c\in\{0, 1\}$ subspace, the evolution of the
QND Hamiltonian for an interaction time of $\tau=\pi/ (2\chi)$
yields \cite{jiang}
\begin{equation} \label{cs}
U  =\exp\left[ -i\tau H_{QND}\right] =\left\{
\begin{tabular}{cc}
$\mathbf{I}$ & ~~~for $n_{c}=0$ \\
$\left( -i\right)^{N}\prod_{l}\sigma_{l}^{z}$ & ~~~for $n_{c}=1$
\end{tabular} \right.
\end{equation}
where $N$ being the number of the selected atoms. If the
cavity is initially prepared in the $n_c=1$ state, the global
operation in Eq. (\ref{cs}) reduces to a string operation
$U_z=\prod_{l}\sigma_{l}^{z}$. Note that all the string operators
are equivalent to $U_z$ up to local single spin rotations
\cite{jiang}: $U_x=\prod_{l}\sigma_{l}^{x}=HU_zH$ and
$U_y=\prod_{l}\sigma_{l}^{y}=RU_zR$, where $H=\prod_{l}H_{l}$ and
$R=\prod_{l}R_{l}$ with
$H_l=\left(\sigma_l^x+\sigma_l^z\right)/\sqrt{2}$ being the
Hadamard rotation and $R_l=\exp\left(-i{\pi \over 4}
\sigma^z_l\right)$. Therefore, as we can implement $U_z$,
universal single qubit gates on  the TSQ can be implemented.

When the cavity state is in a superposition  state of
$\mu|0\rangle_c +\nu|1\rangle_c$, the global operation in Eq.
(\ref{cs}) reduces to  a controlled-string operation:
$U_1=|0\rangle_c\langle0|\otimes I+  |1\rangle_c\langle1|\otimes
U_z$. Here, we need to engineer the cavity photon states. However,
it is actually easier to control an ancillary atom rather than to
directly manipulate the photon number state. Therefore,  we
put an ancillary atom into the optical cavity, which is used to
achieve controlled-string operations between the ancillary atom
and the TSQ. The level structure of the ancillary atom is shown in
Fig. 1(c), the transitions of $|1\rangle\leftrightarrow|e\rangle$
and $|2\rangle\leftrightarrow|e\rangle$ are coupled to the cavity
field with strength $g'$ and a laser with Rabi frequency
$\Omega_A$, respectively. Meanwhile, the two couplings are in a
two-photon resonance scenario with a red detuning $\delta'$ to the
exited state $|e\rangle$. Then, the two couplings are described by
an effective Hamiltonian given as
\begin{equation}
H_e= \lambda
(a^\dagger|2\rangle\langle1|+a|1\rangle\langle2|),
\end{equation}
where $\lambda=g' \Omega_A/\delta'$. To implement the controlled
operations conditioned on the states of the ancillary atom, we
choose the initial state as that the  cavity mode is in a vacuum
state $|0\rangle$ and the ancillary atom is in an arbitrary
superposition state of $\alpha|0\rangle_A + \beta |1\rangle_A$.
The procedure is listed as follows. (1) An interaction of $H_e$
for $t=\pi/\lambda$ coherently couples the cavity  with the atom:
$\alpha|0\rangle_A |0\rangle_c+ \beta |2\rangle_A |1\rangle_c$.
(2) The QND Hamiltonian for $\tau=\pi/ (2\chi)$ on the above
intermediate state is applied. (3) Applying $H_e$ for another
$t=\pi/\lambda$ will annihilate the  cavity photon and restore the
ancillary atom to its original state. In the steps (1) and (3), we
have neglected a phase factor, which can be compensated by a
sing-qubit rotation on the ancillary atom. In this way, one
realizes a controlled-string operation conditioned on the state of
the ancillary atom \cite{jiang}
\begin{equation}
U_2=|0\rangle_A\langle0|\otimes I+  |1\rangle_A\langle1|\otimes U_z.
\end{equation}
In particularly, upon local single spin rotations, one can obtain
the controlled operations  for the TSQ conditioned on the
ancillary atom given as
\begin{eqnarray*}
 U_{cs}^z &=& |0\rangle_A\langle0|\otimes I+
|1\rangle_A\langle1|\otimes S_{\texttt{TSQ}}^z,\\
U_{cs}^x &=& |0\rangle_A\langle0|\otimes I+
|1\rangle_A\langle1|\otimes S_{\texttt{TSQ}}^x,
\end{eqnarray*}
where $S_{\texttt{TSQ}}^{(x,z)}$ are the Pauli matrices for the TSQ.

With such  controlled operations, one is able to access an
efficient TQM \cite{jiang}.  For this purpose, we need a swap in
gate defined as  $U_{in}=H_A U_{cs}^z H_A U_{cs}^x$ with $H_A$
being the Hadamard rotation on the ancillary atom, which swaps the
ancillary atomic state $(\alpha|0\rangle + \beta |1\rangle)_A$
into the topological memory initialized in $|\psi_+\rangle$. For
the inverse process, we need a swap out gate $U_{out}= U_{cs}^x
H_A U_{cs}^z H_A$ that swaps the stored information back to the
ancillary atom prepared in $|0\rangle_A$. The swap in (out)
process  corresponds to write (read) process for our TQM. The read
process also provides us an alternative way of reading out the
topological qubit, which is usually a hard problem.

We now turn to discuss the experimental feasibility of storing the
ancillary  atomic state into the TQM. (1) The TSQ considered here
is robust in the large $N$ limit \cite{kitaev}. For finite $N$,
the ground-state degeneracy is broken, and thus causes decoherence.
However, the lifetime of the TSQ is exponentially increased with
the increasing of $N$. Therefore, for small $N$, one may already
obtain a relatively long coherence time for the TSQ \cite{liu}.
(2) To implement the  QND Hamiltonian, we use the large detuned
scheme, which only requires large Purcell factor, i.e.,
$P=g^2/(\gamma\kappa) >1$ with $\gamma$ and $\kappa$ being the
spontaneous decay rate of level $|e\rangle$ and the cavity decay rate,
respectively. For our large detuned scheme, the effective
spontaneous decay rate is suppressed to $\gamma_{eff}=\gamma g^2 /
\delta^2$. Therefore, as the selected atoms decay independently,
the total probability for photon loss is $P_{loss}^N=(\kappa +
N\gamma_{eff})\tau \geq 2\pi \sqrt{N/P}\approx3\%$ for $N=5$,
$g/(2\pi)= 220$ MHz  \cite{qed}, $\gamma=10$ MHz, and $\kappa=1$
MHz. (3) The addressing errors of the lasers are associated with a
finite spread around the lattice points of atoms, which results in
a tiny coupling between the addressing beam and the selected
atoms, and thus leads to a finite lifetime for the  level
$|2\rangle$. The error probability associated with addressing each
site is estimated to be $\varepsilon=1\%$ \cite{decouple1,decouple2}, which
can  be further suppressed. (4) The deviation of the QND interaction, which degrades the controlled string operation, can also be corrected by the quantum
control techniques to arbitrarily high order \cite{control}. (5)
For the controlled string operations, we further need the
light-atom interface, i.e., the reversible state transfer between
light and the ancillary atom. With a strong lase field of
$\Omega_A=1$ GHz, the error rate can be achieved under
$P_{i}=1\%$. Finally, combining the above error channels, we may
obtain a fidelity about $95\%$ for a controlled operation. As both
the read and write processes of the TQM need two controlled
operations,  a fidelity of $F_1=95\%\times 95\%\approx 90\%$ can
be obtained for the storage of the ancillary atomic state into the
TQM.

\section{Interface with a superconducting qubit}

We now show that by incorporating the above system with an
additional quantum opto-electro-mechanics interface \cite{strong}, we can achieve the storage of a superconducting qubit state into our TQM. The
combined setup is shown in Fig. 1(a), where we consider the case
of two-cavity optomechanical system. For our TQM
purpose, we can firstly transfer the state of the superconducting
qubit to the ancillary atomic qubit and then store the ancillary
atomic qubit state into the TSQ as proposed previously. Therefore,
we only need to consider the state transfer  between the
superconducting qubit and the ancillary atomic qubit in the
following. This process is achieved by the quantum
opto-electro-mechanical setup. In a circuit QED system, under the
rotating-wave approximation, the interaction  Hamiltonian takes the
Jaynes-Cummings form \cite{cqed}
\begin{equation}
H_{m} = g_m \left(b^\dag \sigma^-_s + b \sigma^+ _s \right),
\end{equation}
where $g_m$ is the coupling strength of the superconducting qubit
to the microwave cavity, the subscript  "s" denotes that the
operators belong to the superconducting qubit, $b$ and $b^\dag$
are the annihilation and creation operators of the microwave
cavity field, respectively. Similarly, the ancillary atom is also
coupled to the optical cavity \cite{qed}, as shown in
Fig. 1(c), the transitions of $|1\rangle\leftrightarrow|e\rangle$
and $|0\rangle\leftrightarrow|e\rangle$ are coupled to the cavity
field with strength $g'$ and a laser with Rabi frequency
$\Omega_B$,  in the two-photon resonance way. Then, the two couplings are described by an effective Hamiltonian given as
\begin{equation}
H_o= g_o
(a^\dagger|0\rangle_a\langle1|+a|1\rangle_a\langle0|),
\end{equation}
where $g_o=g' \Omega_B/\delta'$ and the subscript  "a" stands for atom. Here, we consider that the opto-electro-mechanical coupling is enhanced by strongly driving of each cavity, resulting in an effective linear couplings \cite{linear}.
Assuming that each cavity  is far into  the resolved-sideband
regime and  is driven near the red-detuned mechanical sideband, in
the interaction picture, the interaction Hamiltonian reads
\begin{eqnarray}  \label{om}
H _c = G_{1} \left(d a^\dag + d^\dag  a \right)  + G_{2} \left(d b^\dag + d^\dag b \right),
\end{eqnarray}
where $d$  and $d^\dagger$ are the annihilation and creation
operators of  the mechanical oscillator, respectively.  The
coupling between the mechanical resonator and cavity $i$ is
denoted as $G_i$, which is controllable  as they are proportional
to the external driven amplitude, so we may choose $G_1=G_2=G$ for simplicity. The total Hamiltonian reads
$H=H_o+H_m+H_c$, which  conserves the total excitations and we
restrict our discussions to be within the zero- and single-excitation subspaces.
Transfer of the intra-cavity qubit states can be accomplished by
modulating parameters of our system \cite{yin}. As the light-matter
interaction is tunable, we can modulate $g_o=g_m=g$.  The initial
states of the ancillary atomic and superconducting qubits are
assumed to be $|0\rangle_a$ and
$|\psi_s\rangle=(\alpha|0\rangle_s+\beta|1\rangle_s)$,
respectively. Deterministic quantum state transfer is realized
when $\exp{(-itH)}|000\rangle_c |0\rangle_a|\psi_s\rangle
=|000\rangle_c|\psi_a\rangle|0\rangle_s$ is fulfilled, where
$|\psi_a\rangle=\alpha|0\rangle_a+\beta|1\rangle_a$ and
$|000\rangle_c$ means that the three bosonic modes are all in the
vacuum states.

After diagonalizing the total Hamiltonian without dissipation,
we find that deterministic state transfer is achieved at time
$t=\pi/g$, when the relation $2r^2=(4k^2-1)$ (with $k=1, 2, 3,
...$ and $r=G/g$) is fulfilled. To facilitate the state transfer
process,  larger $G$ is preferable \cite{strong}. Meanwhile, to
achieve adiabatic transfer of qubits state, $r\gg 1$ is required
in order to single out only the dark bosonic mode, which results
in much smaller $g$ for a given $G$. As a result, the time needed
to complete the transfer will be much longer, and thus decoherence
will cause considerable  errors. However, if very strong $G$ is
experimentally accessible, e.g., $r=20$, which means $G\sim 100$
MHz, we have numerically confirmed that  the influences of
mechanical mode decay on the state transfer process can be safely
neglected.

\begin{figure}[tb]\centering
\includegraphics[width=8cm]{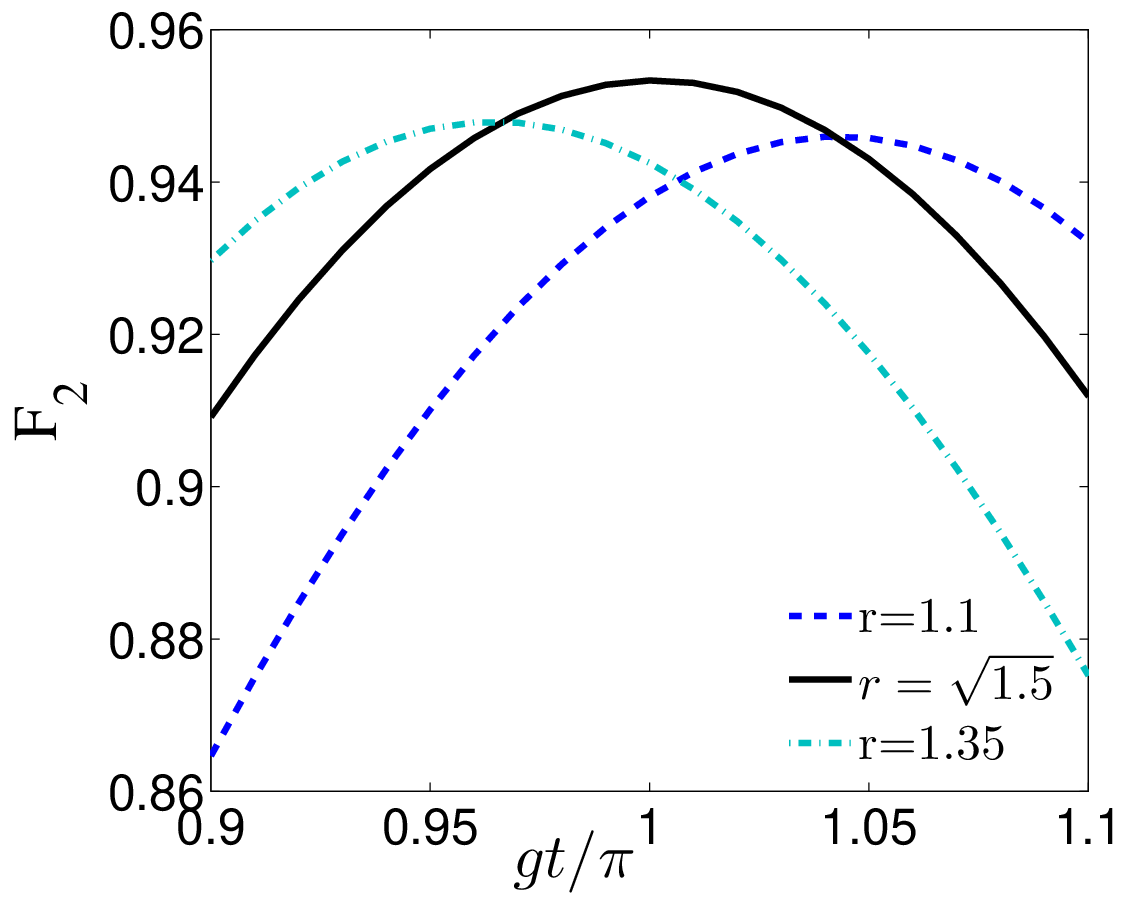}
\caption{The fidelity of the quantum state transfer
between a superconducting qubit and the ancillary atom   as a
function of $gt/\pi$. The parameters are $\kappa_d=0.1$ MHz,
$\kappa_b=\kappa_a=1$ MHz, $\gamma_a=\gamma_s=0.1$ MHz, and $g=6
\pi$ MHz. }  \label{f}
\end{figure}

Finally, we estimate the influence of  dissipation to the state transfer process by  integrating the  quantum master equation
\begin{eqnarray}  \label{me}
\dot{\rho}=-i [H, \rho]
+\sum_\beta \kappa_\beta (2\beta\rho \beta^\dagger - \beta^\dagger\beta \rho -\rho\beta^\dagger \beta )\nonumber\\
+\gamma_s(2\sigma_s^-\rho\sigma_s^+ -\sigma_s^+  \sigma_s^-\rho -\rho\sigma_s^+ \sigma_s^- )\nonumber\\
+\gamma_a(2\sigma_a^-\rho\sigma_a^+ -\sigma_a^+  \sigma_a^-\rho -\rho \sigma_a^+ \sigma_a^-),
\end{eqnarray}
where $\rho$  is the density matrix of the entire system, $\beta
\in \{a, b,d \}$,  $\kappa_a$, $\kappa_b$ and $\kappa_d$  are the
decay rates of the optical  cavity, microwave cavity, and the
mechanical oscillator, respectively; $\gamma_a$ and $\gamma_s$ are
the lifetimes of the atomic and superconducting qubits,
respectively. We characterize the transfer process for the given
initial state by the conditional fidelity of the quantum state
defined by $F_2=\langle\psi_a|\rho_a|\psi_a\rangle$ with $\rho_a$
being the atomic reduced density matrix. By choosing the typical
parameters:  $\kappa_a=1$ MHz, $\kappa_b=1$ MHz, $\kappa_d=0.1$
MHz, $\gamma_a=0.1$ MHz and $\gamma_s=0.1$ MHz, we plot, in Fig.
2, the fidelity of the quantum state transfer process $F_2$ as a
function of the dimensionless time $gt/\pi$, where we have
obtained a high fidelity $F_2>95\%$  of the process. In
particularly, even with considerable deviation (about $\pm10\%$)
of the coupling strength to the ideal condition, $F_2>94\%$ can
still be obtained. Therefore, this process is very robust to the
deviation of the coupling strength. In the above estimation, we
have neglected the effect from the atoms in the optical lattice
due to the following two reasons. Firstly, the cavity-assisted
interaction can be effectively switched off. Secondly, if there is
a small probability that it has not been switched off, then it
will cause energy shift of the cavity mode. As for $N$ atoms, this
shift is $Ng^2/\delta \sim N g/10$ for $\delta\sim 10g$. For the
ancillary atom, we may choose $\delta' =10 \Omega_A =100 g$; in
this way $\lambda=0.1 g$ is still large enough for our
manipulation purpose. With the above parameters, for $N=5$, we
obtain that the atom-induced  energy shift is
$\delta'/200$, and thus can be safely neglected. Therefore, combining
the process of storing the ancillary atomic state into the TQM, we
can obtain a fidelity of $F=F_1\times F_2\approx 86\%$ for storing
a superconducting qubit into the TQM.\\

\section{Conclusion}

In summary, we have proposed a hybrid implementation of a TQM
for both atomic and superconducting qubits, which can combine the
advantages of both the noise resistance of the topological qubits and
the scalability of the superconducting qubits. In particular, by
introducing a quantum opto-electro-mechanical interface, we have demonstrated
that the superconducting qubit state can be efficiently
transferred into the TSQ.

\bigskip
This work was supported by the NFRPC (Grants No. 2013CB921804 and No. 2012CB921604), the NSFC (Grants No. 11474153, No. 11274069, No. 11474064, NO. 61435007, and NO. 11474177), the PCSIRT (Grant No. IRT1243), and the RGC of Hong Kong (HKU173051/14P and HKU173055/15P).


\begin{thebibliography}{99}

\bibitem{s2} You J Q, Nori F. Atomic physics and quantum optics using superconducting circuits. Nature, 2011, \textbf{474}: 589-597

\bibitem{s1} Clarke J, Wilhelm F K. Superconducting quantum bits. Nature, 2008, \textbf{453}: 1031-1042

\bibitem{k03} Kitaev  A. Fault-tolerant quantum computation by anyons.
Ann. Phys. (N.Y.), 2003, \textbf{303}: 2-30

\bibitem{mz} Hasan M Z, Kane  C L. Topological insulators. Rev. Mod. Phys., 2010, {\bf 82}: 3045-3067

\bibitem{p} Qi X L, Zhang S C. Topological insulators and superconductors. Rev. Mod. Phys., 2011, \textbf{83}: 1057-1110


\bibitem{1d} Guo H M. A brief review on one-dimensional topological insulators and superconductors. Sci China-Phys Mech Astron, 2016, \textbf{59}: 637401

\bibitem{t1} Kane C L, Mele E  J. Z$_2$ topological order and the quantum spin Hall effect.  Phys. Rev. Lett., 2005, {\bf 95}: 146802

\bibitem{t2} Bernevig  B A, Hughes T L, Zhang S C. Quantum spin Hall effect and the topological phase transition in HgTe quantum wells.  Science, 2006, {\bf 314}: 1757-1761

\bibitem{t3} Fu  L, Kane C L, Mele E J. Topological insulators in three dimensions. Phys. Rev. Lett., 2007, {\bf 98}: 106803

\bibitem{t4} Moore J  E, Balents L. Topological invariants of time-reversal-invariant band structures. Phys. Rev. B, 2007, {\bf 75}: 121306(R)

\bibitem{t5} Roy R. Topological phases and the quantum spin Hall effect in three dimensions.  Phys. Rev. B,  2009, {\bf 79}: 195322

\bibitem{gauge1} Ruseckas J, Juzeli\={u}nas  G, \"{O}hberg P, Fleischhauer  M. Non-Abelian gauge potentials for ultracold
atoms with degenerate dark states. Phys. Rev. Lett., 2005, {\bf 95}: 010404

\bibitem{gauge2} Zhu S L, Fu H, Wu C J, Zhang S C, Duan L M. Spin Hall effects for cold atoms in a light induced gauge potential. Phys. Rev. Lett., 2006,  \textbf{97}: 240401

\bibitem{gauge3} Zhu S L, Shao L B, Wang Z D, Duan L M. Probing non-abelian statistics of Majorana fermions in ultracold atomic superfluid. Phys. Rev. Lett., 2011, {\bf 106}: 100404

\bibitem{Lin} Lin Y J, Jim\'{e}nez-Garc\'{\i}a K, Spielman I  B. A spin-orbit coupled Bose-Einstein condensate. Nature, 2011, {\bf 471}: 83-86

\bibitem{Wang} Wang P, Yu Z Q, Fu Z, et al. 
   Spin-orbit coupled degenerate Fermi gases. Phys. Rev. Lett., 2012, {\bf 109}: 095301


\bibitem{MIT} Cheuk L W, Sommer A T, Hadzibabic Z, Yefsah T, Bakr W S, Zwierlein   M W. Spin-injection spectroscopy of a spin-orbit coupled Fermi gas. Phys. Rev. Lett., 2012, {\bf 109}: 095302

\bibitem{Pan} Zhang J Y, Ji S C, Chen Z, et al. 
    Collective dipole oscillation of a spin-orbit coupled Bose-Einstein condensate. Phys. Rev. Lett., 2012, {\bf 109}: 115301


\bibitem{simulation} Bloch I, Dalibard J, Zwerger W. Many-body physics with ultracold gases. Rev. Mod. Phys., 2008, \textbf{80}: 885-964


\bibitem{liu}  Liu X J, Liu Z X, Cheng M. Manipulating  topological edge spins in ome-dimensional optical lattice. Phys. Rev. Lett., 2013, \textbf{110}: 076401

\bibitem{jiang} Jiang L, Brennen G K, Gorshkov A V, et al. Anyonic interferometry and protected memories in atomic spin lattices. 
Nat. Phys., 2008, \textbf{4}: 482-488

\bibitem{Wu2015}  Wu Z,  Zhang L, Sun W, et al. Realization of two-dimensional spin-orbit coupling for Bose-Einstein condensates. arXiv:1512.06394

\bibitem{h1}  Kubo Y, Grezes C, Dewes A, et al. Hybrid quantum circuit with a superconducting qubit coupled to a spin ensemble. Phys. Rev. Lett., 2011, \textbf{107}, 220501

\bibitem{h2}     Yang W L, Yin Z Q, Hu Y, Feng M, Du J F. High-fidelity quantum memory using nitrogen-vacancy center ensemble for hybrid quantum computation. Phys. Rev. A, 2011,  \textbf{84}: 010301

\bibitem{h3}  Chen Q, Yang W L, Feng M. Controllable quantum state transfer and entanglement generation between distant nitrogen-vacancy-center ensembles coupled to superconducting flux qubits.   Phys. Rev. A, 2012, \textbf{86}: 022327

 \bibitem{h4} Tao M J, Hua M, Ai Q, Deng F G. Qauntum information processing on nitrogen-vacancy ensembles with the local resonance assisted by circuit QED.  Phys. Rev. A, 2015, \textbf{91}: 062325

\bibitem{strong} Bagci T, Simonsen A, Schmid S, et al.  
Optical detection of radio waves through a nanomechanical tansducer.
Nature, 2014, \textbf{507}: 81-85

\bibitem{interface}  Andrews R W, Peterson R W, Purdy T P, et al. Bidirectional and efficient conversion between microwave and optical light.
    Nat. Phys., 2014, \textbf{10}: 321-326
\bibitem{Yin2015} Yin Z Q, Yang W L, Sun L, Duan L M. Quantum network of superconducting qubits through opto-mechanical interface, Phys. Rev. A, 2015, \textbf{91}: 012333

\bibitem{om} Li T, Yin Z Q. Quantum superposition, entanglement, and state teleportation of a microorganism on an electromechanical oscillator. Science Bulletin, 2016 \textbf{2}: 163-171

\bibitem{linear} Xiong H, Si L G, Lv X Y, et al. %
Review of cavity optomechanics in the weak-coupling regime: from linearization to intrinsic nonlinear interactions. Sci China-Phys Mech Astron, 2015, 58: 50302


\bibitem{cqed} Schoelkopf R J, Girvin S M. Wiring up quantum systems. Nature, 2008, \textbf{451}, 664-669

\bibitem{inter1} Jiang L, Kane C L, Preskill J.
Interface between topological and superconducting qubits.
Phys. Rev. Lett., 2011, \textbf{106}: 130504


\bibitem{inter2} Xue Z Y, Shao L B, Hu Y, Zhu  S L, Wang Z D.
Tunable interfaces for realizing universal quantum computation with topological qubits. Phys. Rev. A, 2013, \textbf{88}: 024303


\bibitem{inter3} Cottet A, Kontos T, Dou\c{c}ot B.
Squeezing light with Majorana fermions.
Phys. Rev. B, 2013, \textbf{88}: 195415


\bibitem{inter4} Xue Z Y, Gong M, Liu J, Hu Y, Zhu  S L, Wang Z D.
Robust interface between flying and topological qubits. Sci. Rep., 2015, \textbf{5}: 12233



\bibitem{kitaev} Kitaev A Y. Unpaired Majorana fermions in quantum wires. Physics-Uspekhi, 2001, \textbf{44}: 131-136



\bibitem{decouple1} Cho J. Addressing individual atoms in optical lattices standing-wave driving fields. Phys. Rev. Lett., 2007, \textbf{99}: 020502


\bibitem{decouple2} Gorshkov A V, Jiang L, Greiner M, Zoller P, Lukin M D. Conherent quantum optical control with subwavelength resolution. Phys. Rev. Lett., 2008, \textbf{100}: 093005


\bibitem{qed} Hood C J, Lynn T W, Doherty A C, Parkins A S, Kimble  H J. The atom-cavity microscope: Single atoms bound in orbit by single photons.  Science, 2000, \textbf{287}: 1447-1453

\bibitem{control} Brown K R,  Harrow A W,  Chuang I  L. Arbitrarily accurate composite pulses. Phys. Rev. A, 2004, \textbf{70}: 052318


\bibitem{yin} Yin Z Q, Li F L. Multiatom and resonant interaction scheme for quantum state transfer and logical gates between two remote cavities via an optical fiber. Phys. Rev. A, 2007, \textbf{75}: 012324



\end{thebibliography}
\end{document}